\documentclass[12pt]{article}
\usepackage{graphicx}
\usepackage{amssymb}
\usepackage{bm}
\evensidemargin \oddsidemargin
\def\0{{\bf 0}}
\def \E {{\cal E}}
\def \LL {{\cal L}}

\def\b#1{{\mathbb #1}}

\newcommand{\bx}{{\bf x}}

\newcommand{\bxp}{{\bf x}^{\scriptscriptstyle \perp}}
\newcommand{\hbxp}{\hat {\bf x}^{\scriptscriptstyle \perp}}
\newcommand{\bXp}{{\bf X}^{\scriptscriptstyle \perp}}
\newcommand{\bX}{{\bf X}}
\newcommand{\bu}{{\bf u}}
\newcommand{\bup}{{\bf u}^{\scriptscriptstyle \perp}}
\newcommand{\hbup}{\hat {\bf u}^{\scriptscriptstyle \perp}}

\newcommand{\hbu}{\widehat{\bf u}}
\newcommand{\bw}{{\bm w}}

\newcommand{\bv}{{\bf v}}

\newcommand{\rx}{{\rm x}}
\newcommand{\Ba}{{\bm \alpha}}
\newcommand{\Bap}{{\bm \alpha}^{\scriptscriptstyle \perp}}
\newcommand{\bb}{{\bm \beta}}

\newcommand{\Be}{{\bm \epsilon}}
\newcommand{\Bep}{{\bm \epsilon}^{\scriptscriptstyle \perp}}
\newcommand{\bE}{{\bf E}}
\newcommand{\bEp}{{\bf E}^{\scriptscriptstyle \perp}}
\newcommand{\hbEp}{\hat{\bf E}^{\scriptscriptstyle \perp}}

\newcommand{\bB}{{\bf B}}
\newcommand{\bBp}{{\bf B}^{\scriptscriptstyle \perp}}
\newcommand{\hbBp}{\hat{\bf B}^{\scriptscriptstyle \perp}}

\newcommand{\bK}{{\bf K}}
\newcommand{\bKp}{{\bf K}^{\scriptscriptstyle \perp}}
\newcommand{\bA}{{\bf A}}
\newcommand{\bAp}{{\bf A}^{\scriptscriptstyle \perp}}

\newcommand{\bjp}{{\bf j}^{\scriptscriptstyle \perp}}

\newcommand{\Bp}{{\bf p}}
\newcommand{\bP}{{\bf P}}
\newcommand{\bPi}{{\bf \Pi}}

\newcommand{\bi}{\mathbf{i}}
\newcommand{\bj}{\mathbf{j}}
\newcommand{\bk}{\mathbf{k}}

\newcommand{\be}{\begin{equation}}
\newcommand{\ee}{\end{equation}}
\newcommand{\bea}{\begin{eqnarray}}
\newcommand{\eea}{\end{eqnarray}}
\newcommand{\ba}{\begin{array}}
\newcommand{\ea}{\end{array}}
\begin{document}

\title{Travelling waves and light-front approach in relativistic electrodynamics
}


\author{Gaetano Fiore$^{1,2}$,        Paolo Catelan$^{3,4}$ \\  \\
$^{1}$ Dip. di Matematica e Applicazioni, Universit\`a ``Federico II'',\\
   Complesso MSA, V. Cintia, 80126 Napoli, Italy;\\         
$^{2}$  I.N.F.N., Sezione di Napoli, \\
        Complesso MSA, V. Cintia, 80126 Napoli, Italy; \\
$^{3}$  
Centro de Energ\'ias Alternativas y Ambiente, \\
Escuela Superior Polit\'ecnica del Chimborazo, Riobamba, Ecuador;\\
$^{4}$  Dip. di Matematica ed Informatica, 
Universit\'a della Calabria, \\ Arcavacata, Rende, Italy
}


\date{}

\maketitle

\begin{abstract}
We briefly report on a recent proposal \cite{Fio18JPA} for simplifying the 
equations of motion of  charged particles in an electromagnetic  (EM) field $F^{\mu\nu}$ that is the sum of
a plane travelling wave \ $F_t^{\mu\nu}(ct\!-\!z)$ \ and a static part \ $F_s^{\mu\nu}(x,y,z)$; \
it  adopts the light-like coordinate $\xi=ct\!-\!z$ instead of time $t$ as an independent variable. 
We illustrate it  in a  few  cases of extreme acceleration, first of an isolated particle, then of electrons
in a plasma in plane hydrodynamic conditions: 
the Lorentz-Maxwell \& continuity PDEs can be simplified or sometimes 
even completely reduced to a family of decoupled systems of ordinary ones;
this occurs e.g. with the impact of the travelling wave on a vacuum-plasma interface
(what may produce plasma waves or the  slingshot effect). 



\end{abstract}

\section{Introduction}
\label{intro}

The equation of motion of a charged particle in an external EM  field  
\bea
\ba{l}
\displaystyle\dot\Bp(t)=q\bE[t,\bx(t)] + \frac{\Bp(t) }{\sqrt{m^2c^2\!+\!\Bp^2(t)}}  \wedge q\bB[t,\bx(t)] 
,\\[6pt]  
\displaystyle
\dot \bx(t) =\frac{c\Bp(t) }{\sqrt{m^2c^2\!+\!\Bp^2(t)}}
\ea
\label{EOM}
\eea
in its  general  form is non-autonomous and highly nonlinear in the unknowns $\bx(t),\Bp(t)$. \
Here \ $m,q,\bx,\Bp$ \ are the  rest mass, electric charge, position
and   relativistic momentum of the particle,  \ $\bE=-\partial_t\bA/c-\nabla A^0$ 
and $\bB=\nabla\!\wedge\!\bA$ are the electric and magnetic field, $(A^\mu)=(A^0,-\bA)$ is the electromagnetic (EM) potential 4-vector ($E^i=F^{i0}$, $B^1=F^{32}$, etc.; we use  Gauss CGS units).
We decompose $\bx\!=\!x\bi\!+\!y\bj\!+\!z\bk\!=\bxp\!+\!z\bk$, etc, in the cartesian coordinates of the laboratory frame, and often use the dimensionless variables \ $\bb\!\equiv\!\bv/c\!=\!\dot\bx/c$,  
\  $\gamma\!\equiv\! 1/\sqrt{1\!-\!\bb^2}\!=\!\sqrt{1\!+\! \bu^2}$ \ and
the 4-velocity $u\!=\!(u^0\!,\bu)\!\equiv\!(\gamma,\gamma \bb)
$, i.e. the dimensionless version of the 4-momentum $p$.
Usually, (\ref{EOM}) is  simplified assuming:

\begin{enumerate}

\item $\bE,\bB$ are constant or vary  ``slowly'' in space/time; \ or

\item $\bE,\bB$  are  ``small" (so that nonlinear effects  in $\bE,\bB$ are negligible);  \ or

\item $\bE,\bB$ are monochromatic waves, or slow modulations of;   \ or

\item the  motion of the particle keeps non-relativistic.

\end{enumerate}

The on-going, astonishing developments of  laser technologies  today allow the construction of compact
sources of extremely intense coherent EM waves,
possibly concentrated in very short laser pulses. {\it Chirped Pulse
Amplification}  \cite{StriMou85,MouTajBul06}  allows the production of pulses of intensity
up to $10^{23}$ Watt per square centimeter and duration down to $ 10^{-15}$ seconds.
Huge investments in  new technologies (thin film compression,  relativistic
mirror compression, etc.  \cite{MouMirKhaSer14,TajNakMou17}) 
will soon allow to produce even more intense/short (or cheaper) 
pulses. For instance,  850 MEuro have been allocated for the 
{\it Extreme Light Infrastructure} (ELI)  program within the European Union ESFRI roadmap, with
three of the planned four sites already under construction in Czech Republic, Hungary, Romania. 
One major motivation is the quest for  table-top particle accelerators based on Laser Wake Field Acceleration (LWFA)
\cite{Tajima-Dawson1979} in plasmas. Among the possible applications of such accelerators we mention:

\begin{itemize}

\item  {\bf Medicine}: inspection (PET,...), cancer therapy by accelerated particles (electrons, protons, ions) or radioisotope production,...;

\item {\bf Research}: particle physics, materials science, structural biology, (inertial) nuclear  fusion,
X-ray free electron laser,...;

\item {\bf Industry}: atomic scale lithography, surface treatment of materials, sterilization, energy efficient 
manufacturing, detection systems,...;

\item {\bf Environmental remediation}: flue gas cleanup, petroleum cracking, transmutation of nuclear wastes,....

\end{itemize} 
These and other applications of small accelerators were discussed e.g. at the 
``Big Idea Summit'' organized by the US Department of Energy (Washington, 2016).
In Europe the large network of research centers ``European Plasma Research Accelerator with eXcellence 
In Applications" (EUPRAXIA) has been recently created to develop the associated technologies.
Extremely intense and rapidly varying electromagnetic fields 
are present also in several violent astrophysical processes
(see e.g. \cite{TajNakMou17} and references therein).
In either case the effects are so fast, huge, highly nonlinear, ultra-relativistic
that conditions 1-4  are not fulfilled. Alternative simplifying approaches are
therefore welcome.

\begin{figure}
\includegraphics[width=15cm]{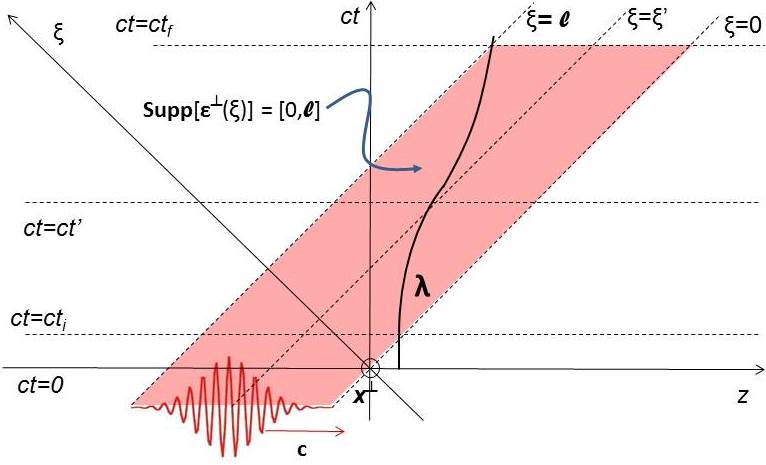}
\caption{Every worldline $\lambda$  and hyperplane $\xi\!=\xi'$
in Minkowski space intersect once. The wave-particle interaction
occurs only along the intersection of $\lambda$ with the support of the EM wave (painted pink), which 
assuming (\ref{aa'}) is delimited by the  $\xi\!=\!0$ and $\xi\!=\!l$ hyperplanes.}
\label{Worldline}       
\end{figure}

\smallskip
Here we summarize an approach \cite{Fio18JPA} that systematically
applies the light-front formalism \cite{Dir49}; it is especially fruitful if in the spacetime region 
of interest
(where we wish to follow the particles' worldlines) $\bE,\bB$  are the sum of
static parts and plane transverse travelling waves propagating in the $z$ direction:
\be
\bE(t,\bx)=\!\!\underbrace{\Bep(ct\!-\!z)}_{ travelling\, wave}\!+\underbrace{\bE_s(\bx)}_{static},\qquad \bB(t,\bx)=\underbrace{\bk\wedge\Bep(ct\!-\!z)}_{travelling\, wave} +
\underbrace{\bB_s(\bx)}_{static}.
\label{EBfields}
\ee
The starting point is:  as no particle can 
reach the speed of light $c$, then $\tilde \xi(t)\!=\!ct\!-\!z(t)$ is strictly growing and  
we can make the change  $t\mapsto \xi\!=\!ct\!-\!z$ of independent parameter  along the worldline  
$\lambda$ (fig. \ref{Worldline}) of the particle; then the term $\Bep[ct\!-\!z(t)]$, where the unknown $z(t)$ is in the argument of the highly nonlinear and rapidly varying $\Bep$, becomes the known forcing term $\Bep(\xi)$. 
We apply the approach first to an isolated particle (sections \ref{GenRes}, \ref{Exact}), then to a cold diluted plasma initially at rest and hit by a plane EM wave  (section \ref{Plasmas}).

The fields (\ref{EBfields}) can be obtained from an EM potential of the same form, \
$A^\mu(\rx)=\alpha^\mu(ct\!-\!z)+A_s^\mu(\bx)$; \
in the Landau gauges ($\partial_\mu A^\mu\!=\!0$) \ $\bA_s$ must fufill the
Coulomb gauges  ($\nabla\!\cdot\! \bA_s\!=\!0$), and it must be \ $\alpha^z{}'=\alpha^0{}'$, 
 \ $\Bep\!=\!-\Bap{}'$, $\bE_s\!=\!-\nabla\! A_s^0$, $\bB_s\!=\!\nabla \!\wedge\! \bA_s$. \
We shall set  $\alpha^z=\alpha^0=0$, as they  appear neither in the observables $\bE,\bB$ 
\ nor in the equations of motion.
Assuming only   that $\Bep(\xi)$ is piecewise continuous and
\bea
\ba{ll}
&\mbox{ {\bf a}) }\quad \Bep \mbox{ has a compact support }[0,l],\\[10pt] 
\mbox{ or} &\mbox{ {\bf a'})}\quad  \Bep \in L^1(\mathbb{R}),
 \ea   \label{aa'} 
\eea
we can fix $\Bap(\xi)$ uniquely by requiring that
it vanish as $\xi\to-\infty$:
\bea
\Bap(\xi)\!\equiv\!-\!\!\int^{\xi}_{ -\infty }\!\!\!\!\!\!\!dy\,\Bep\!(y);         \label{defBap}
\eea
in case {\bf a}) 
$\Bap(\xi)\!=\!0$ if $\xi\!\le\! 0$,  $\Bap(\xi)\!=\!\Bap(l)$ if $\xi\!\ge\! l$. \
We can treat on the same footing all $\Bep$ fulfilling (\ref{aa'}) regardless of their Fourier analysis, in particular:
\begin{enumerate}
\item A modulated monochromatic wave:
\be
\Bep\!(\xi)\!=\!\underbrace{\epsilon(\xi)}_{\mbox{modulation}}
\underbrace{[\bi a_1\cos (k\xi\!+\!\varphi)\!+\!\bj a_2\sin (k\xi)]}_{\mbox{carrier wave $\Be_o^{{\scriptscriptstyle \perp}}\!(\xi)$}}
 \label{modulate}
\ee
(with $a_1^2\!+\!a_2^2\!=\!1$). Under rather general assumptions
\be
\Bap(\xi)= -   \frac {\epsilon(\xi)}k \:\Bep_p\!(\xi)+O\left(\frac 1 {k^2}\right)
 \simeq -   \frac {\epsilon(\xi)}k \,\Bep_p\!(\xi),              \label{slowmodappr}
\ee
$\Bep_p\!(\xi)\! :=\!\Bep_o(\xi\!+\!\pi/2k)$; in the appendix we  recall upper bounds  for the remainder $O(1/k^2)$.
For slow modulations  (i.e. $|\epsilon'|\!\ll\! |k\epsilon|$) 
 - like the ones characterizing most conventional applications (radio broadcasting, ordinary laser beams,  etc.)  -
the right estimate is  very good.
\label{modula1}

\item 
A superposition of waves of type 1.
\label{modula2}

\item An `impulse'  (few, one, or even a fraction of oscillation) \cite{Aki96,
MouMirKhaSer14}.

\end{enumerate}

\section{Set-up and general results for a single particle}
\label{GenRes}

Let $\hat \bx(\xi)$ be the position  as a function of $\xi$; 
it is determined  by $\hat \bx(\xi)=\bx(t)$. More generally,  for any given function $f(t,\bx)$ we 
denote $\hat f(\xi, \hat \bx)\equiv f[(\xi\!+\!  \hat z)/c,  \hat \bx]$,
 abbreviate $\dot f\!\equiv\! df/dt$, $\hat f'\!\equiv\! d\hat f/d\xi$ (total derivatives). 
Also the change of dependent (and unknown) variable $u^z\mapsto s$ is convenient, where the {\it $s$-factor}  \cite{Fio18JPA}
\be
s\equiv\gamma\!- u^z=u^-=\gamma(1-\beta^z)=\frac{\gamma}c \frac{d\tilde \xi}{dt}>0          \label{defs0}
\ee
is the light-like component of $u$ (the dimensionless version of $p$), 
as well as the Doppler factor of the particle. In fact,
$\gamma,\bu,\bb$ are {\it rational}  functions of $\bu^{{\scriptscriptstyle\perp}}, s$:
\be
\gamma\!=\!\frac {1\!+\!\bu^{{\scriptscriptstyle\perp}}{}^2\!\!+\!s^2}{2s}, 
\qquad  u^z\!=\!\frac {1\!+\!\bu^{{\scriptscriptstyle\perp}}{}^2\!\!-\!s^2}{2s}, 
 \qquad  \bb\!=\! \frac{\bu}{\gamma}                                    \label{u_es_e}
\ee
(these relations hold also with the carets); so, replacing $d/dt\mapsto(c s/ \gamma)d/d\xi$ and putting carets  on all variables (\ref{EOM}) becomes {\it rational} in the unknowns $\hat\bu^{{\scriptscriptstyle\perp}},\hat s$: 
\bea
\ba{l}
 \displaystyle\hat\bx'=\frac{\hat\bu}{\hat s},\qquad\quad
\hat\bup{}' =\frac q{mc^2\, \hat s}\!\left[\hat\gamma\hat\bE\!+\!\hat\bu\!\wedge\!\hat\bB\right]^{\scriptscriptstyle \perp}\!
, \\[12pt]
 \displaystyle \hat s'  = \frac {q}{mc^2}\!\left[\frac{\hat\bup}{\hat s}\!\cdot\!\hat\bEp\!-\!\hat E^z\!-\!\frac{(\hat\bup\!\wedge\!\hat\bBp)^z}{\hat s}\right]
\ea
\label{equps0}
\eea
with $\hat u^z,\hat \gamma$ expressed as  in (\ref{u_es_e}). 
\ These equations amount  \cite{Fio18JPA}
to the  Euler-Lagrange equations 
$\frac d {d\xi}\frac{\partial \LL}{\partial \hat \bx'}=\frac{\partial \LL}{\partial \hat \bx}$,  \  that are obtained
applying Hamilton's principle to the action functional $S(\lambda)$ with $\lambda$ parametrized by $\xi$ (instead of $t$), as well as 
to the Hamilton equations \
$\hat \bx'\!=\!\frac{\partial \hat H}{ \partial \hat\bPi},\: \hat\bPi' \!=\!-\frac{\partial \hat H}{ \partial \hat \bx}$, where the Hamiltonian 
\bea
\hat H(\hat \bx,\!\hat\bPi;\!\xi)={mc^2}\frac {1\!+\! \hat s^2\!\!+\! \hat \bu^{{\scriptscriptstyle\perp}2}\!}{2\hat s}
\! +\!q\hat A^0\!(\xi,\hat\bx),
\quad \mbox{with}\:\left\{\!\!\ba{l} \displaystyle\hat\bu^{{\scriptscriptstyle\perp}}\!\!=\!\frac{\hat\bPi^{{\scriptscriptstyle\perp}}\!
\!-\!q\hat \bA^{{\scriptscriptstyle\perp}}(\xi,\hat\bx)}{mc^2} \\[6pt]
\displaystyle\hat s\!=\!-\frac{\hat \Pi^z\!\!+\!q[\hat A^0\!\!-\!\hat A^z](\xi,\hat\bx)}{mc^2},        
\ea\right.     \label{Ham}
\eea
 is obtained by Legendre transform from $\LL$ and again is rational in 
$\hat\bPi\equiv\frac{\partial \LL}{\partial \hat \bx'}$,
or equivalently in $\hat\bu^{{\scriptscriptstyle\perp}},\hat s$. Along the  solutions
$\hat H$ gives the particle energy  as a function of $\xi$, and
\be
\frac {d \hat H} {d\xi}=\frac {\partial \hat H} {\partial\xi}.
\label{derH}
\ee
Under the EM field  (\ref{EBfields}) equations  (\ref{equps0}) amount to 
\bea
\hat\bx'=\frac{\hat\bu}{\hat s},\qquad\quad\ba{l}\displaystyle\hbup{}'\!=\frac q{mc^2}\!\left[(1\!+\!\hat z')\hbEp_s\!+\!(\hat\bx'\!\wedge\!\hat\bB_s)^{\scriptscriptstyle \perp}\!+\!\Bep(\xi) \right]\!,\\[10pt]
\displaystyle\hat s'=\frac {-q}{mc^2}\left[\hat E^z_s-\hbxp{}'\!\cdot\!\hbEp_s\!+(\hbxp{}'\!\wedge\!\hbBp_s)^z\right]\!,    
\ea\qquad \label{equps} 
\eea
while the  {\it energy gain}  (normalized to the rest energy $mc^2$) in the interval $[\xi_0,\xi_1]$ is 
\be
\E:= \frac{\hat H(\xi_1)\!-\!\hat H(\xi_0)}{mc^2}= \int^{\xi_1}_{\xi_0}\!\!\! d\xi\:\frac{q\Bep \!\cdot\hbup}{mc^2\hat s }(\xi).
\label{EnergyGain}
\ee 
In particular, under assumption (\ref{aa'}a) we obtain the total energy gain choosing \ $\xi_0\!=\!0$, $\xi_1\!=\!l$, \
which are the values of the lightlike coordinate at the beginning and at the end of the interaction, see fig. \ref{Worldline}.
If we used parameter $t$, to compute $\E$ we should first determine the time $t_f$
when the pulse-particle interaction fineshes. 
Once solved (\ref{equps}), analytically or numerically, to obtain the solution as a function of $t$ we just need to invert $\hat t (\xi)\!=\!\xi\!+\!\hat z(\xi)$ and set $\bx(t)=\hat\bx[\xi(t)]$.

Contrary to  (\ref{equps}),  (\ref{EOM}) is not rational in $\bu$, and
the unknown $z(t)$ appears
in the argument of the rapidly varying functions $\Bep,\Bap$ in (\ref{EOM}a), which now reads:
\bea
\frac {mc}q \dot\bu(t)\!=\! \bE_s\!\!+\!\frac {\bu\!\wedge\!\bB_s} {\sqrt{1\!+\!\bu^2}}\!+\!\frac {\bu\!\cdot\!\Bep[ct\!-\!z(t)]} {\sqrt{1\!+\!\bu^2}}\bk\!+\!\left(1\!-\!\frac {u^z} {\sqrt{1\!+\!\bu^2}}\right)\Bep[ct\!-\!z(t)].
\nonumber 
\eea
$H(\bx,\bP,t)$ is not rational in $\bP\!:=\!\frac{\partial L}{\partial \dot x}$, and also determining $\E(t)$ is more complicated.

\subsection{Dynamics under a $A^\mu$ independent of the transverse coordinates}
\label{xiz}

Eq. (\ref{equps0})  are further simplified if  \  $A^\mu\!=\!A^\mu(t,\!z)$.  \ \ 
This applies in particular  if $\bE_s= E_s^z(z)\bk$, $\bB_s=\bB_s^{\scriptscriptstyle \perp}(z)$, choosing \ $A^0=-\int^z\!d\zeta E_s^z(\zeta)$, 
\ $\bAp\!=\Bap\!-\!\bk\wedge\!\int^z\!d\zeta \bBp(\zeta)$, $A^z\!\equiv\! 0$.
As $\partial \hat H/\partial \hat \bx^{\scriptscriptstyle \perp}\!=\!0$, we find
$\hat\bPi^{\scriptscriptstyle \perp}\!=\!q\bKp\!=\!\mbox{const}$, i.e.
the known result $\frac {mc^2}q\hat\bu^{\scriptscriptstyle \perp}\!=\!\bKp
\!-\!\hat \bA^{{\!\scriptscriptstyle\perp}}\!(\xi,\!\hat z)$.
  Setting $v\!:=\!\hbup{}^2$  and replacing  in   (\ref{equps}) we obtain 
\bea
\hat z'=\displaystyle\frac {1\!+\!\hat v}{2\hat s^2}\!-\!\frac 12, \qquad
\hat s'=\frac{-q}{mc^2}E_s^z(\hat z)\!-\!\frac 1{2\hat s}
\frac{\partial \hat v}{\partial \hat z}.     \label{reduced}
\eea
Once solved system (\ref{reduced}) for $\hat z(\xi), \hat s(\xi)$,
the other unknowns are  obtained from 
\bea
&& \hat\bx(\xi)=\bx_0+\!\displaystyle\int^\xi_{\xi_0}\!\!\! dy \,\frac{\hat\bu(y)}{\hat s(y)}.\qquad \label{hatsol}
\eea
\smallskip
[the $z$-component of (\ref{hatsol}) amounts to  (\ref{reduced}a)  with initial condition $\hat z(\xi_0)\!=\!z_0$]. 
If in addition $\bB_s\!\equiv\!0$, then 
 $\bA_s\!\equiv\!0$, implying that
\ $\hat\bu^{\scriptscriptstyle \perp}\!(\xi)\!=\!\frac q{mc^2}\left[\bK^{\scriptscriptstyle \perp}\!-\Bap(\xi)\right]$ and 
$ \hat v\!=\!\hat\bu^{{\scriptscriptstyle\perp}2}$
 are already known. The system   (\ref{reduced}) to  be solved simplifies to
\bea
&& \hat z'=\frac {1\!+\! \hat v}{2\hat s^2}\!-\!\frac 12, \qquad  \hat s'=\frac{-q}{mc^2}E_s^z(\hat z).
 \label{heq1r} 
\eea
{\bf Remarks.} Some remarkables properties of the corresponding solutions are  \cite{Fio18JPA}:  

\begin{enumerate}

\item Where $\Bep(\xi)\!=\! 0$ then $ \hat v(\xi)\!=\!v_c\!=$const,
$\hat H$ is conserved,  (\ref{heq1r}) is solved by quadrature.

\item 
\label{notransv}
In case (\ref{aa'}a) the final transverse momentum is \ $mc\hat\bup(l)$.
 If  $\epsilon$ of (\ref{modulate}) varies slowly and $\hat\bup\!(0)\!=\!\0$, then by (\ref{slowmodappr}) 
$\hat\bup(l)\!\simeq\!0$.

\item 
Fast  oscillations of $\Bep$ make $\hat z(\xi)$ oscillate much less than $\hat\bxp(\xi)$, and 
$\hat s(\xi)$ even less: as $\hat s\!>\!0$, $\hat v\!=\!\hbup{}^2\!\ge\! 0$, 
integrating (\ref{heq1r}a) averages the fast oscillations of $\bup$ to yield much smaller relative 
oscillations of $\hat z$,  while integrating (\ref{heq1r}b) averages the residual small 
oscillations of $E_s^z[\hat z(\xi)]$ to yield an essentially smooth $\hat s(\xi)$.
On the contrary, $\hat \gamma(\xi), \hat\bb(\xi), \hat\bu(\xi),...$, which 
are recovered via (\ref{u_es_e}), oscillate fast, and so do also $\gamma(t), \bb(t), \bu(t),..$. See e.g. fig. \ref{Ez=0},\ref{Ez=const>},\ref{graphs}.
\label{insensitive}

\item If  $\bup\!(0)\!=\!\0$ and the EM wave is a slowly modulated (\ref{modulate})-(\ref{aa'}a),
integrating   (\ref{EnergyGain}) by parts across $[0,l]$ and using  (\ref{slowmodappr}) we find
\ $\E\simeq\int^{l}_{0}\! d\xi \, \hat  v(\xi)\hat s'(\xi)/2\hat s^2(\xi)$: \ the 
 energy gain will be automatically positive (resp. negative) if $\hat s(\xi)$ is growing 
(resp. decreasing) in all of $[0,l]$. Correspondingly, the  interaction with the EM wave can
be used to accelerate (resp. decelerate) the particle. 
\label{signEnGain}

\end{enumerate}

\section{Some solutions in closed form under constant  $\bB_s,\bE_s$}
\label{Exact}

Assume $\bB_s,\bE_s$ are constants, and let ${\bf b}\!:=\! q\bB_s/mc^2$, 
${\bf e}\!\equiv\! q\bE_s/mc^2$.
Upon integration over $\xi$ and use of (\ref{equps}a) equations (\ref{equps}b-c) yield
\bea
\ba{l}
\hat u^x= (e^x\!-\!b^y)\hat z\!+\!b^z \hat y\!+\! w^x(\xi), \\[6pt]
\hat u^y= (e^y\!+\!b^x)\hat z\!-\!b^z  \hat x\!+\! w^y(\xi), \\[6pt]
\hat s=(e^x\!-\!b^y)\hat x \!+\!(e^y\!+\!b^x)\hat y\!-\!w^z(\xi),
\ea    \label{constEsBs'}
\eea
where $\bw(\xi)\! \equiv\! q\!\left[\bK\!\!-\!\Bap(\xi)\!+\xi\bE_s\right]\!/\!mc^2$ ($\bw$ is known and dimensionless),
and $\bK$ is an integration constant.
For any $E^z_s,B^z_s,\bEp_s$, if  $\bBp_s\!=\!\bk\wedge\bEp_s$,  then $e^x\!=\!b^y$,
$e^y\!=\!-b^x$, and (\ref{constEsBs'}c)  is solved: $\hat s\!=\!-w^z(\xi)$.
Then we solve  in closed form the rest of the system (\ref{constEsBs'}), (\ref{equps}a) first for $\hat\bxp(\xi)$, 
then for $\hat\bup(\xi), \hat u^z(\xi),\hat z(\xi)$. 
Assuming for simplicity the initial conditions $\bx(0)\!=\!\0\!=\!\bu(0)$  we find
 \bea
\ba{l}
\displaystyle ( \hat x +i \hat y)(\xi)=(1\!-\!{\rm e}^z\xi)^{ib^z/{\rm e}^z}\int^\xi_0\!\!\!\!  d\zeta \,
 \frac{(w^x+iw^y) (\zeta)} {(1\!-\!{\rm e}^z\zeta)^{1+ib^z/{\rm e}^z}}, \\[12pt]
\displaystyle   \hat z(\xi)\!=\!\!\int\limits^\xi_0\!\!  \frac {d\zeta }{2}\!
\left[\frac {1}{(1\!-\!{\rm e}^z\zeta)^2}\!+\!\hbxp{}'{}^2\!(\zeta)\!-\!1\right]\!, \quad
\hat s(\xi)\!=\!1\!-\!{\rm e}^z\xi, \\[12pt]
 \hbup(\xi)\!=\!(1\!-\!{\rm e}^z\xi)\,\hbxp{}'(\xi),  \qquad
  \hat \gamma(\xi)\!=1\!-\!{\rm e}^z\xi\!+\!\hat u^z(\xi)                      \\[14pt]
\displaystyle            \hat u^z(\xi)\!=\!\frac {1}{2(1\!-\!{\rm e}^z\xi)}+
(1\!-\!{\rm e}^z\xi)\,\frac {\hbxp{}'{}^2(\xi)\!-\!1}{2}
\ea   \label{SolEqBzEz}
\eea
if ${\rm e}^z\neq 0$ \ and
\bea
\ba{l}
\displaystyle  ( \hat x \!+\! i \hat y)(\xi)=\int^\xi_0\!\!\!\!  d\zeta \, e^{-ib(\xi\!-\!\zeta)}(w^x\!\!+\! iw^y) (\zeta),
 \qquad \hat\bup\!=\!\hat \bxp{}',\\[12pt]
\displaystyle  \hat u^z\!=\!\hat z'\!=\!\frac {\hat\bup{}^2}{2}=\E(\xi)=\hat\gamma(\xi)-1, \qquad
\hat z(\xi)\!=\!\int^\xi_0\!\!\!\!  d\zeta \:\frac {\hat\bup{}^2(\zeta)}{2}.
\ea   \label{SolEqBz}
\eea
if ${\rm e}^z= 0$. \ As fas as we now, such general  solutions have not appeared 
in the literature before Ref. \cite{Fio18JPA}. 
We next analyze a few special cases (the first two have already appeared  in the literature).

\subsection{Case $\bE_s\!=\!\bB_s\!=\!\0$ (zero static fields). \ Then (\ref{SolEqBzEz}) becomes \cite{LanLif62,Fio14JPA}:}
\label{LW}

\bea
\ba{l}
\displaystyle\hat s\!\equiv\! 1, \qquad  \hat\bu^{\scriptscriptstyle \perp}\!\!=\!
\frac {-q\Bap}{mc^2}, \qquad \hat u^z\!=\!\frac {\hat  \bup{}^2}2, \quad\hat\gamma\!=\!1\!+\!\hat u^z\\[16pt] 
\displaystyle \hat z(\xi)\!=\! \int ^\xi_{0}\!\!\!\!dy\,\frac{\hat\bu^{{\scriptscriptstyle\perp}2}(y)}2,\qquad
\quad \hbxp\!(\xi)\!=\! \int ^\xi_{0}\!\!\!\!dy\,\hbup\!(y).      \ea                      \label{U=0s=0}
\eea
The solutions (\ref{U=0s=0}) induced by two $x$-polarized pulses 
and the corresponding electron trajectories  in the $zx$ plane  are shown in fig. \ref{Ez=0}. Note that:
\begin{figure*}
\centering
\includegraphics[width=7.8cm]{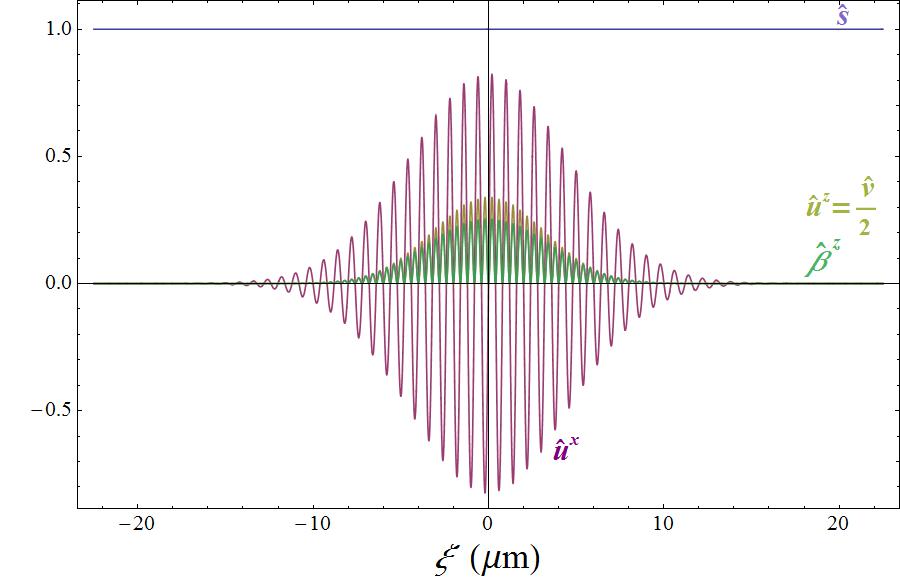} \hfill  \includegraphics[width=7.8cm]{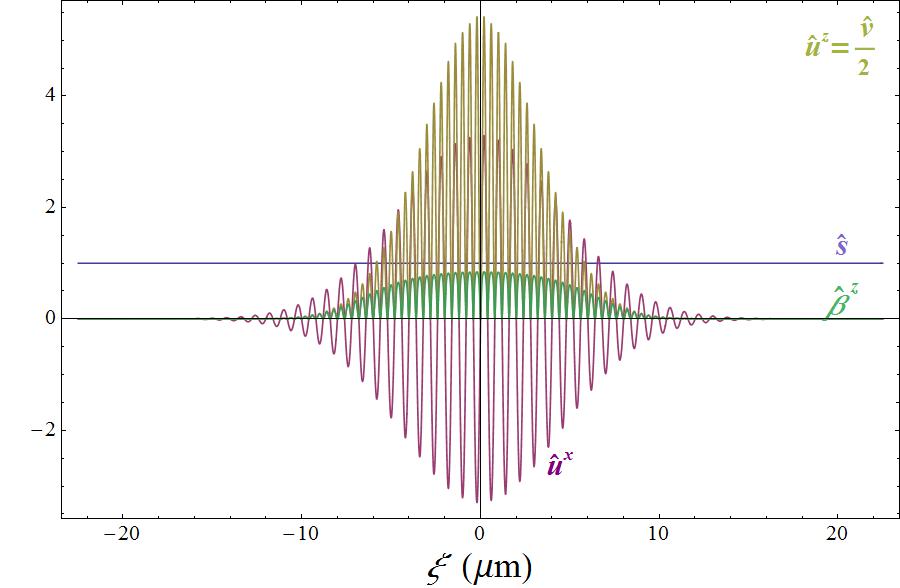}\\
\includegraphics[width=7.8cm]{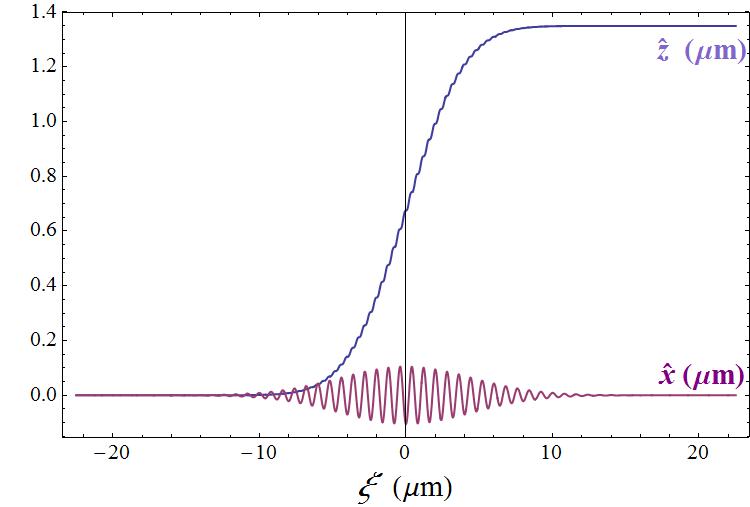}\hfill \includegraphics[width=7.8cm]{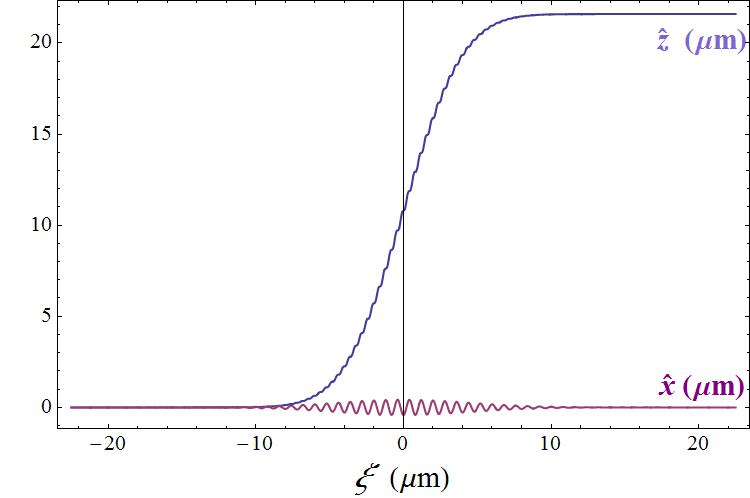} \\
\vskip.3cm
\includegraphics[width=5.3cm]{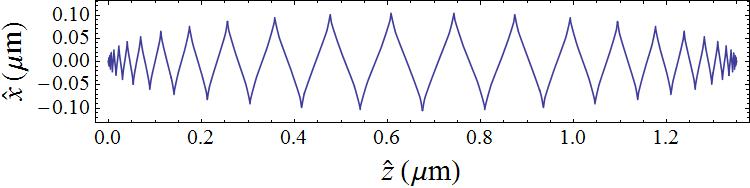} \hfill 
\includegraphics[width=10cm]{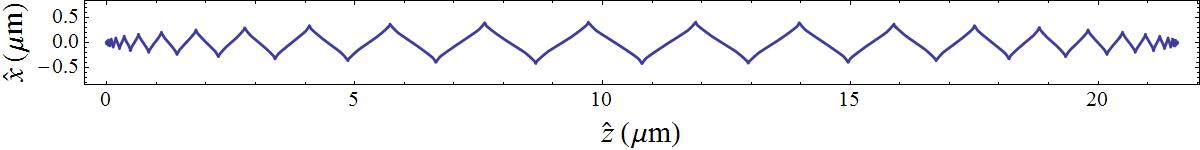}
\caption{Solutions (\ref{U=0s=0}) (up and center) and corresponding electron trajectories 
 in the $zx$ plane (down) induced  by two $x$-polarized pulses with 
carrier wavelength $\lambda\!=\!.8\mu$m, gaussian modulation $\epsilon(\xi)=a\exp[-\xi^2/2\sigma]$,
$ \sigma \!=\! 20\mu$m$^2$, \ $|q| a\lambda/mc^2  \!=\! 4, 15$ (left, right). 
}
\label{Ez=0}      
\end{figure*}

$\bullet$ \ The maxima of $\gamma$, $\alpha^{\scriptscriptstyle \perp}$ 
coincide (and approximately also of $\epsilon(\xi)$, if $\epsilon(\xi)$ is slowly varying).

\smallskip
$\bullet$ \ Since $u^z\!\ge\!0$,  the $z$-drift is nonnegative-definite.
If we rescale $\Bep\mapsto a\Bep$ then $\hbxp,\hat \bup$ scale like $a$, 
whereas  $\hat z,\hat u^z$  scale like $a^2$;
hence the trajectory goes to a straight line in the limit $a\!\to\!\infty$. 
This is due to  magnetic force $q\bb \wedge \bB$.

\smallskip
$\bullet$  \ {\bf Corollary}  \cite{Fio18JPA} \
The final $\bu$ and energy gain  read
\be
\bup_f\!=\!\hbup(\infty), \qquad u^z_f=\E_f= \frac 12\bup_f{}^2=\gamma_f\!-\! 1
 \label{Lawson-1} 
\ee
[in case (\ref{aa'}a) it is also $\bup_f\!=\!\hbup(l)$].
By (\ref{slowmodappr}), both are very small if the pulse modulation $\epsilon$ is slow [extremely small if 
\ $\epsilon\!\in\!{\cal S}(\mathbb{R})$ \ or \ $\epsilon\!\in\!C^\infty_c(\mathbb{R})$].
This can be seen as a rigouros version of the {\it Lawson-Woodward}  
Theorem \cite{Law84,Pal88,
Pal95,EsaSprKra95} 
 (an outgrowth of the original Woodward-Lawson Theorem  \cite{Woo47,WooLaw48}): 
this theorem states that, 
 in spite of large energy variations during the interaction, 
the final energy gain $\E_f$ of a charged particle  interacting
with an EM field is zero if: 

i)  the interaction occurs in $\b{R}^3$ vacuum (no boundaries); 

 ii)  $\bE_s=\bB_s=\0$  and $\Bep$ is slowly modulated;  

iii)  $v^z\simeq c$ along the whole acceleration path; 

 iv) nonlinear  (in $\Bep$) effects $q\bb\!\wedge\! \bB$ are negligible; 

v) the power radiated by the particle is negligible. 

\noindent 
Our Corollary, as Ref. \cite{TrohaEtAl99}, states that the same result holds
if we relax iii), iv),
but the EM field is a {\it plane} travelling wave. 
\ To obtain a non-zero $\E_f$ one has to violate some other  conditions 
of the theorem, as e.g. we consider in next cases.

\begin{figure}[ht]
\includegraphics[width=7.8cm]{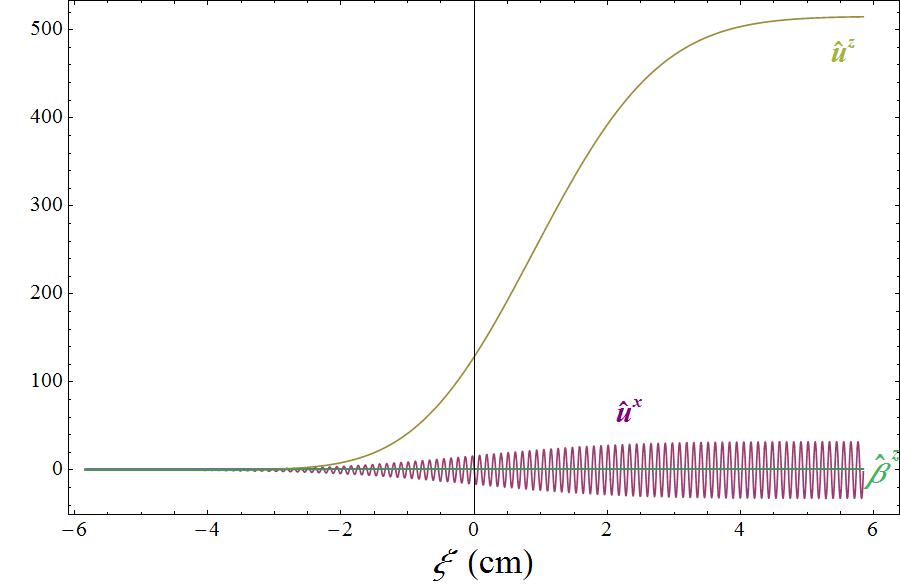} \hfill \includegraphics[width=7.8cm]{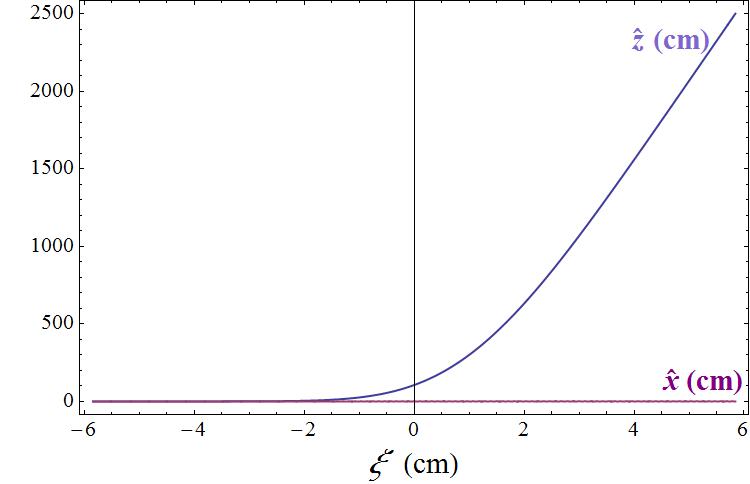}\\[12pt]
\includegraphics[width=16.4cm]{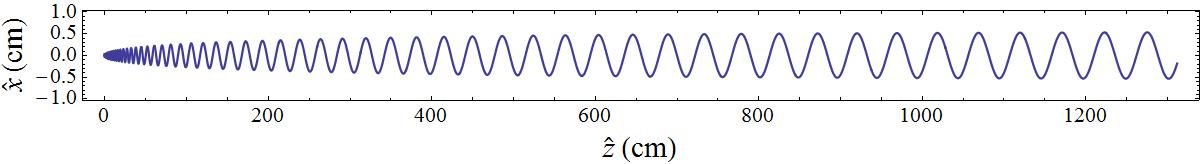}
\caption{The electron 
motion  (\ref{SolEqBz}) (up) and the $zx$-projection of the corresponding trajectory (down) induced 
in a longitudinal magnetic field $B^z\!=\! 10^5$G by a circularly polarized modulated EM wave  (\ref{modulate}) with wavelength $\lambda\!\equiv\!2\pi/k\!=\!1$mm, 
$b \!=\! k=\!58.6$cm$^{-1}$, gaussian enveloping amplitude $\epsilon(\xi)=a\exp[-\xi^2/2\sigma]$ with
$ \sigma \!=\! 3$cm$^2$ 
and \ $e a/kmc^2\!=\!0.15$,  trivial initial conditions ($\bx_0 \!=\!  \bu_0 \!=\! 0$),  giving $\E_f\!\simeq\! 28.5$.}
\label{Bz=const}      
\end{figure}

\subsection{Case \ $\bE_s=0$, \ $\bB_s=B^z_s\bk$. \ Then the solution (\ref{SolEqBzEz}) becomes
(see fig. \ref{Bz=const})}

\bea
\ba{l}
\displaystyle 
 ( \hat x \!+\! i \hat y)(\xi)\!=\!\!\int^\xi_0\!\!\!\!  d\zeta \, e^{ib(\zeta\!-\!\xi)}(w^x\!\!+\! iw^y) (\!\zeta\!),
 \qquad\quad \hbup\!=\!\hbxp{}'\!,\\[12pt]
\displaystyle  \hat s\equiv 1,\quad\hat u^z\!=\!\hat z'\!=\!\frac {\hbup{}^2}{2}\!=\!\E\!=\!\hat\gamma\!-\!1, \quad
\hat z(\xi)\!=\!\!\int^\xi_0\!\!\!\!  d\zeta \:\frac {\hbup{}^2\!(\zeta)}{2}.
\ea   \label{SolEqBz}
\eea
For monochromatic $\Bep$ it reduces to the solution of \cite{KolLeb63-1,KolLeb63-2,Dav63} 
and leads to {\it cyclotron autoresonance} if 
$-b\!=\!k\!\gg\!\frac 1l$: assuming
for simplicity circular polarization [$a_1\!=\!a_2\!=\!1$ in (\ref{modulate})], 
by (\ref{slowmodappr}) it is \ $w^x(\xi)\!+\!iw^y(\xi)\!\simeq\!  e^{ik\xi} {\rm w}(\xi)$,  whence
\bea
( \hat x \!+\! i \hat y)(\xi)\simeq  i  W\!(\xi)e^{ik\xi}, \quad \hat z(\xi)\simeq 
 \!\! \int^\xi_0\!\!\!\!  d\zeta   \frac {k^2W^2(\zeta)}2  ,  \qquad 
W(\xi)\!:=\!\!\int^\xi_0\!\!\!\!  d\zeta \frac{q\epsilon(\zeta)}{kmc^2}\!>\!0;   
\nonumber \label{ApprSolEqBz-2}
\eea
clearly $W(\xi)$ grows with $\xi$. 
In particular if $\Bep(\xi)\!=\!\0$ for $\xi\!\ge\! l$, then for such $\xi$
$$
\hat z'(\xi)\!\simeq\!  \frac {k^2}2 W^2(l)\!\simeq\!2\E_f, \qquad
\frac{|\hbxp{}'(\xi)|}{\hat z'(\xi)}\!\simeq\!\frac 2 {k W\!(l)}\!\ll\! 1; 
$$
the final energy gain is noteworthy  by the first formula,
the final collimation is very good by the second.

\begin{figure}[ht]
\begin{minipage}{.47\textwidth}
\includegraphics[width=7.3cm]{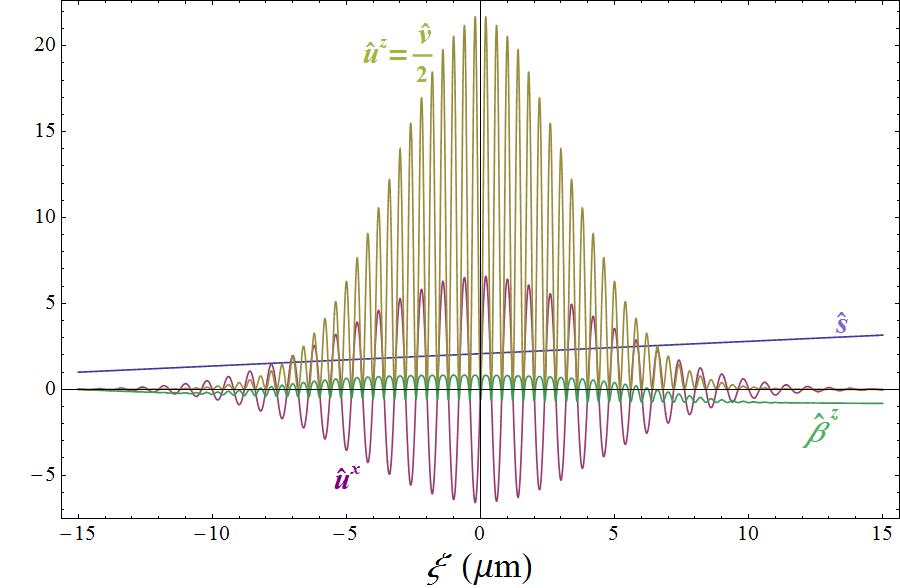}
\\ \includegraphics[width=7.3cm]{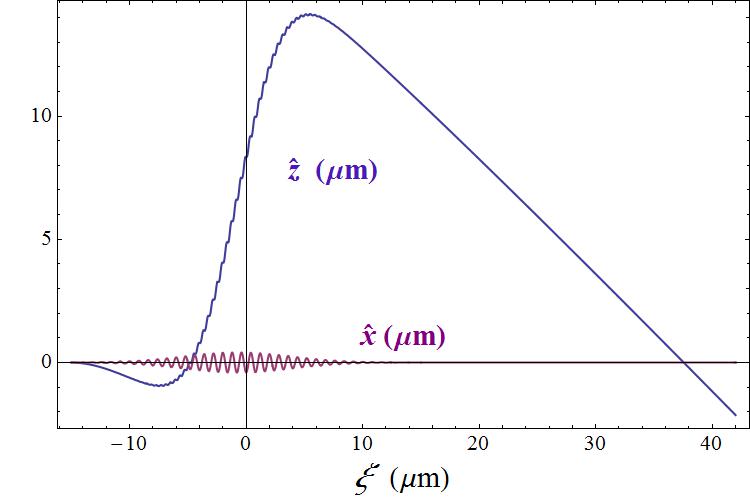}
\end{minipage}%
\hfill\begin{minipage}{.52\textwidth}
\includegraphics[width=8.4cm]{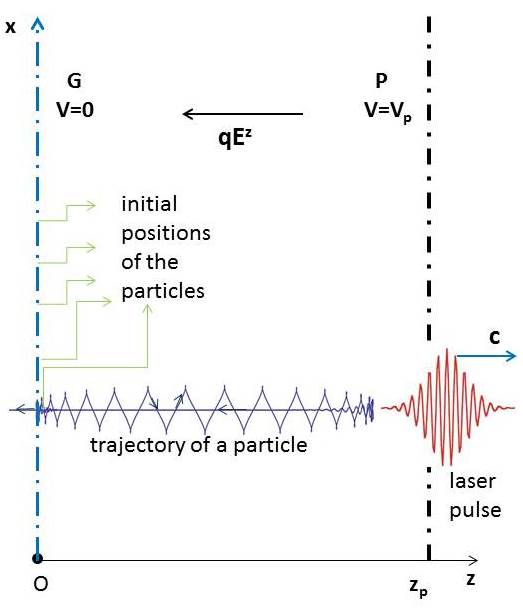}
\end{minipage}
\caption{Left: the motion (\ref{Ezcost})   induced by a linearly polarized modulated EM wave  (\ref{modulate}) with wavelength $\lambda\!=\!2\pi/k\!=\!0.8\mu$m, gaussian enveloping amplitude $\epsilon(\xi)\!=\! a\exp[-\xi^2/2\sigma]$ with
$ \sigma \!=\! 20\mu$m$^2$ 
and $|q|a\sqrt{2}/kmc^2\!=\!6.6$,  trivial initial conditions, $\bB_s \!=\!\0$,
$\bE_s \!=\!\bk E^z_{\scriptscriptstyle M}$, where $E^z_{\scriptscriptstyle M}q\!\simeq\!  37$GeV/m
(this yields the maximum energy gain $\E_f\!\simeq\! 1.5$ with such a wave). Right: the corresponding trajectory in the $zx$ plane within an hypothetical acceleration device based on a laser pulse and 
metallic gratings $G,P$ at potentials $V\!=\!0,V_p$, with $qV_p/z_p\!\simeq\! 37$GeV/m. 
}
\label{Ez=const>}
\end{figure}

\subsection{\ Case \ $\bE_s=E_s^z\bk$, \ $\bB_s=\0$. \ 
The solution (\ref{SolEqBzEz}) reduces to  $\hat s(\xi)=1\!-\! {\rm e}^z \xi $,}
\label{Ezconst}
\be
 ( \hat x +i \hat y)(\xi)=\!\!\int^\xi_0\!\!\!  dy \,
 \frac{(w^x\!\!+\!iw^y) (y)} {1\!-\!{\rm e}^z y},\qquad
\hat z(\xi)=\!\!
\int^\xi_{0}\! \frac {dy}2 \left\{\!\frac {1\!+\! \hat v(y)}{[1\!-\!{\rm e}^z y]^2}\!-\!1\right\}\!;
\label{Ezcost}
\ee
by Remark   \ref{GenRes}.\ref{signEnGain}, 
if $\Bep$ is slowly modulated the energy gain  $\E_f$ 
is negative if  ${\rm e}^z\!\equiv\! qE_s^z/mc^2\!>\!0$, is positive if   ${\rm e}^z\!<\!0$,
and has a unique maximum at some point ${\rm e}^z_{\scriptscriptstyle M}\!<\!0$ if $\epsilon(\xi)$ 
fulfills (\ref{aa'}a) with
a unique maximum. An acceleration device based on this solution would consist of the following: 
at $t\!=\!0$ the particle  is initially at rest with $z_0\!\lesssim\! 0$, just at the left of a metallic grating $G$ contained in the  $z\!=\!0$ plane and set at zero electric  potential (see fig. \ref{Ez=const>});
another metallic plate $P$  contained in a  plane $z\!=\!z_p\!>\!0$  is set at electric potential $V=V_p$.
A short laser pulse $\Bep$ travelling in the positive $z$-direction hits and boosts the particle into the latter region  (see section \ref{LW}); 
choosing $qV_p\!>\!0$ implies ${\rm e}^z\!<\!0$, and a 
backward longitudinal electric force $qE^z_s$. If  $qV_p$ is large enough, then $z(t)$ reaches a maximum
smaller than $z_p$, then is accelerated backwards and exits the grating
with energy $\E_f$ and negligible transverse momentum. 
A large $\E_f$ requires extremely large $|V_p|$, far beyond the material breakdown threshold,
what prevents its realization by a static potential (sparks between $G,P$ would arise and rapidly reduce $|V_p|$).
A way out is to make the pulse itself generate such large  $|E_s^z|$ within a plasma just
at the right time, so as to induce
the {\it slingshot effect}, as sketchily explained at the end of next section.

\begin{figure}[ht]
\includegraphics[width=7.8cm]{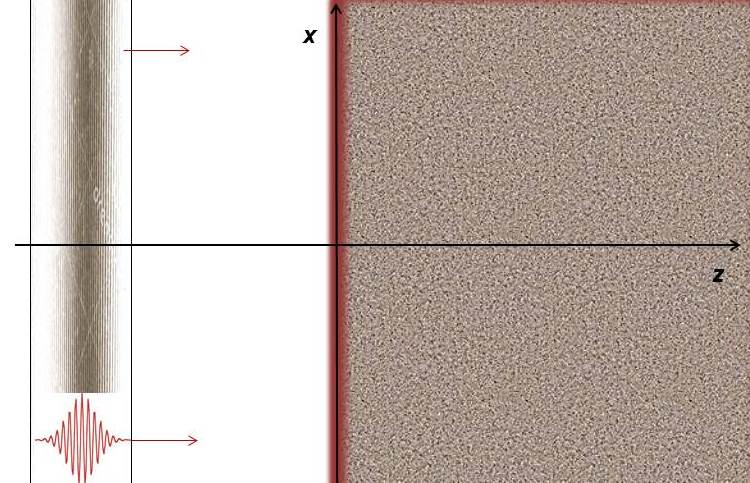}\hfill\includegraphics[width=7.8cm]{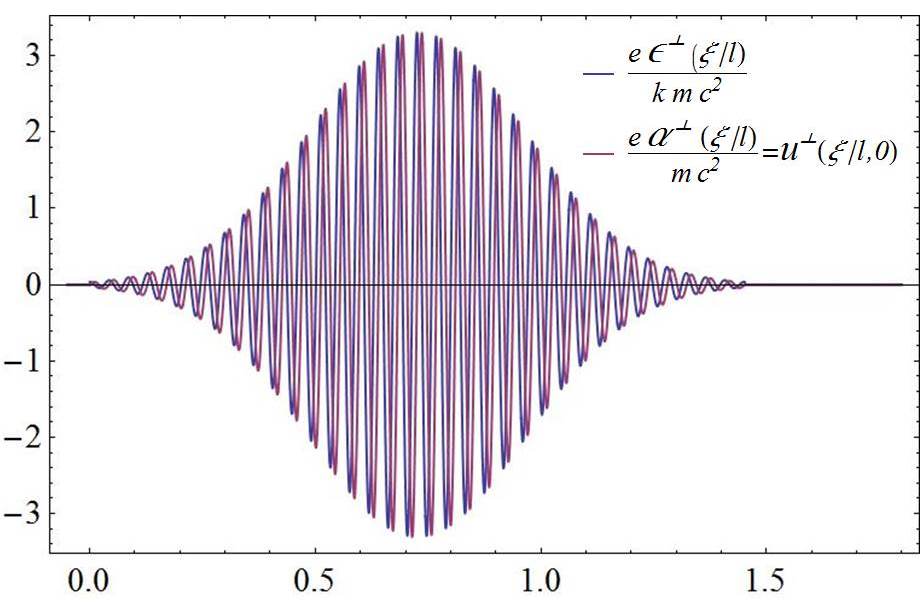} 

\caption{Left: \ a plane EM wave of finite length approaching normally a plasma in equilibrium. \
Right:  \  normalized EM wave $\Bep$ (blue)  with carrier wavelength $\lambda\!=\!0.8\mu$m, linear polarization, 
gaussian modulation $\epsilon(\xi)=a\exp[-\xi^2/2\sigma]$ with
$ \sigma \!=\! 20\mu$m$^2$, \ $e a\lambda/mc^2\!=\!15$ (whence the average pulse intensity is
$10^{19}$W/cm$^2$); the associated $\bup_e$ is painted purple.  $l$ is the length of the $z$-interval  where the amplitude $\epsilon$ overcomes all ionization thresholds  of the atoms of the  gas  yielding the plasma;
here we have chosen helium, whence  $l\!\simeq\! 27\mu$m, and the thresholds for $1^{st}$ and $2^{nd}$ ionization  are overcome  almost simultaneously.
}
\label{Plane}
\end{figure}

\section{Plane plasma problems}
\label{Plasmas}

Assume that the plasma is initially in hydrodynamic conditions with all initial  data [Eulerian velocities $\bv_h$ and densities $n_h$ of the  $h$-th fluid, EM fields of the form (\ref{EBfields}); $h$ enumerates  electrons and kinds of ions composing the plasma, $q_h,m_h$ are their charge, mass] not depending on $\bxp$. Then also the solutions of
the Lorentz-Maxwell  and continuity equations for  $\bB,\bE,\bu_h,n_h$  do not depend on $\bxp$, nor the 
displacements $\Delta\bx_h\equiv \bx_h(t,\bX)\!-\!\bX$ on $\bXp$. \ Here $\bx_h( t,\!\bX)$ is the position at $t$
of the material element of the  $h$-th  fluid  with initial position $\bX\!\equiv\!(X,\!Y,\!Z)$; $\bX_h( t,\bx)$
is the inverse of $\bx_h( t,\!\bX)$ (at fixed $t$); $\bb_h\!=\!\bv_h/c$, etc.
More specifically, we consider  (fig. \ref{Plane}) a very short and intense EM plane wave 
(\ref{aa'}a) hitting normally a cold  plasma  (or a gas that is locally ionized into a plasma by the very high electric field of the pulse itself)  initially in equilibrium, possibly in a static and uniform magnetic field $\bB_s$; 
the initial conditions are:
\bea 
\ba{l}
n_h(0,\bx)\!=\!0\quad \mbox{if }\: z\!\le\! 0,\qquad \bu_h(0,\bx)\!=\!\0,\qquad
 j^0(0,\bx)\!=\!\sum\limits_h\!q_hn_h(0,\bx)  \!\equiv\! 0, \\[8pt]
\bE(0 ,\bx)\!=\!\Bep(-z),\qquad\quad \bB(0,\bx)\!=\!\bk\wedge\Bep(-z) +\bB_s,
\ea                                                       \label{incond0}
\eea
whence the 4-current density $j\!=\!(j^0,\bj)\!=\!\sum\limits_hq_h n_h(1,\bb_h)$
is zero at $t=0$.
Then the Maxwell equations \ $\nabla\!\cdot\!\bE\!=\!4\pi j^0$,
$ \partial_t E^z\!/c +\!4\pi j^z \!\!=\!(\!\nabla\!\!\wedge\!\bB)^z\!\!=\!0$ imply  \cite{Fio14JPA}
\be
E^{{\scriptscriptstyle z}}(t,\!z)=4\pi \sum_h q_h
\widetilde{N}_h[ Z_h(t ,z)], \qquad \widetilde{N}_h(Z)\!:=\!\int^{Z}_0\!\!d \zeta\,n_h(0,\!\zeta);
         \label{expl}
\ee
using (\ref{expl}) to express $E^z$ in terms of the (still unknown) longitudinal motion 
[$ Z_h(t ,\cdot)$  is the inverse of $z_h(t,\cdot)$] we  reduce the number of unknowns by one. 

Define $\Bap$ as in (\ref{defBap}); $\Bap(\xi)\!=\!\0$ if $\xi\!\le\!0$.
\ In the Landau gauges  (\ref{incond0}) are compatible 
with the following initial conditions for the gauge potential: 
\bea
 \bA(0 ,\bx)=\Bap(-z)\!+\!\bB_s\!\wedge\!\bx/2,  
\qquad \partial_t \bA(0, \bx)=-c\Bep(-z),
\label{asyc'}
\eea 
$A^0(0,\!\bx)=\partial_t A^0(0,\!\bx)=0$. \ By (\ref{incond0}-\ref{asyc'})  and causality \ $\bx_h(t,\bX)\!=\!\bX$, 
$\bAp(t ,\bx)\!\equiv\! \bB_s\!\wedge\!\bx/2$ if  $ct\!\le\! z$,    $\bj\!\equiv\!\0$ if  $ct\!\le\! |z|$.  
$\bAp$ is coupled to the current through $\Box\bAp=4\pi\bjp$. 
Including (\ref{asyc'}) the latter amounts to the integral equation
\be
\bA\!^{{\scriptscriptstyle\perp}}\!\!-\!
\Ba\!^{{\scriptscriptstyle\perp}}\!-\!\frac 12\left(\bB_s\!\wedge\!\bx\right)^\perp \!=\!
2\pi \!\! \int \!\!\!  d\!s   d\zeta\, \theta(ct\!\!-\!\!s\!-\!|z\!\!-\!\!\zeta|)
\theta(\!s\!) \,\bjp\!\left(\frac sc,\zeta\right);  
       \label{inteq1}
\ee
here we have used  the Green function of the d'Alembertian $\partial_t^2/c^2\!-\!\partial_z^2$ \ in dimension 2. 
The right-hand side (rhs) is zero for $t\le 0$ ($t=0$ is the beginning  of the laser-plasma interaction). 
Within {\it short} time intervals $[0,t']$ (to be determined {\it a posteriori})  we can thus:
approximate \ $\bAp(t,z)\simeq \Bap(ct\!-\!z)\!+\!\left(\frac {\bB_s}2\!\wedge\!\bx\right)^\perp$; \ 
also neglect the motion of  ions with respect to the motion  of the (much lighter) electrons. Hence it is \ $z_p(t,Z)\!\equiv\!Z$,
\ and the proton density $n_p$  (due to ions of all kinds) equals the initial one and therefore  the initial  electron density $\widetilde{n_0}(z)\!:=\!n_e(0,z)$, by the initial electric neutrality of the plasma.
Then the   equations (\ref{equps0}) \& initial conditions for the electron fluid amount to
\be
\ba{l}
m c^2\hat s_e'(\xi,Z)=4\pi e^2 \left[
\widetilde{N}( \hat z_e ) \!-\! \widetilde{N}( Z)\right] + e(\Delta\hbxp_e{}'\!\wedge\!\hbBp_s)^z\!, \\[10pt]
m c^2 \hbup_e{}'(\xi,Z) =e\Bap{}' - e(\Delta\hat\bx'_e\!\wedge\!\hat\bB_s)^{\scriptscriptstyle \perp}\!,\qquad
\Delta\hat\bx'_e=\frac{\hbu_e(\xi,Z)}{\hat s_e(\xi,Z)}
\ea      \label{equps1}
\ee
\be
\Delta\hat\bx_e(0,\!\bX)\!=\!0, \qquad \hat\bu_e(0,\!\bX)\!=\!\0\qquad\Rightarrow \quad\hat s_e(0,\!\bX)\!=\!1.     
\label{incond}
\ee
\smallskip
(\ref{equps1}) is a family parametrized by $Z$ of {\it decoupled ODEs} in the unknowns
\ $\Delta\hat\bx_e,\hat s_e$, $\hbup_e$, \ 
which can be solved numerically.
The approximation on $\bAp\!(t,\!z)$ is acceptable as long as the so determined motion makes \ $|\mbox{rhs}(\ref{inteq1})| \!\ll\! |\Bap\! + \frac {\bB_s}2\! \wedge \bx|$; \ otherwise rhs(\ref{inteq1}) determines the first correction to $\bAp$; and so on.

\begin{figure}[ht]
.\hskip0.3cm\includegraphics[width=15.9cm]{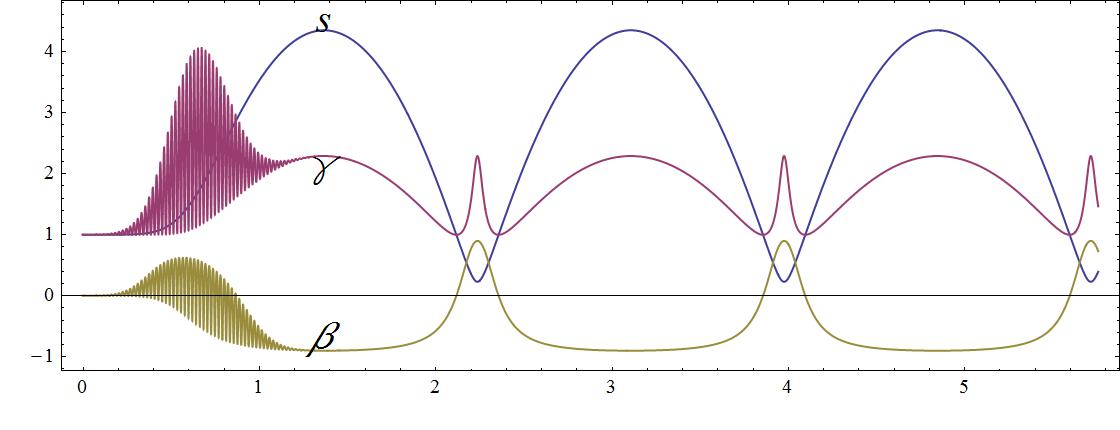} \\
\includegraphics[width=16.4cm]{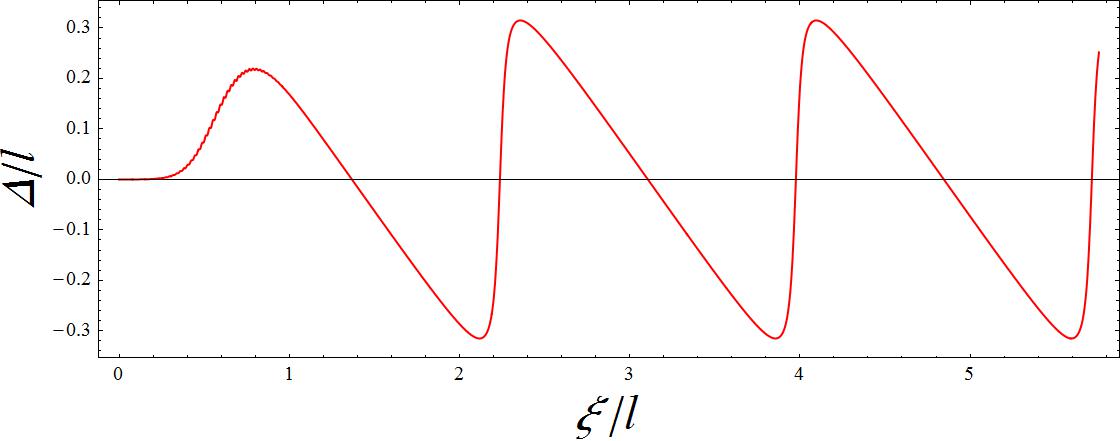} 
\caption{
Down: Solution of (\ref{heq1}-\ref{u_es_e}) corresponding to the pulse of fig. \ref{Plane}-right,
and to the initial electron density \ $\widetilde{n_{0}}(Z)\!=\!n_0\theta(Z)$, \ with  $n_0\!=\! 2 \!\times\! 10^{18}$cm$^{-3}$.   
}
\label{graphs}
\end{figure}

If $\bB_s\!=\!\0$, again (\ref{equps1}b)  is solved by $\hbup_e(\xi)\!=\! e\Bap(\xi)/mc^2$,  
while, setting $ v\!=\!\hat\bu^{{\scriptscriptstyle\perp}2}$, \  (\ref{equps1}a)  and the
$z$-component of (\ref{equps1}c) take   \cite{FioDeN16,FioDeN16b} the form of (\ref{heq1r}),
\bea
\Delta\hat z_e'\!=\displaystyle\frac {1\!+\! v}{2\hat s^2}\!-\!\frac 12, \quad
\hat s'_e\!=\frac{4\pi e^2}{mc^2} \left\{ 
\widetilde{N}[\hat z_e] \!-\! \widetilde{N}(Z) \right\}.  \label{heq1} 
\eea
If $n_e(0,\!\bX)\!=\!n_0\theta(Z)$ (with a constant electron density  $n_0$), then as long as 
$\hat z_e(\xi,Z)\!>\!0$ 
(\ref{heq1}), (\ref{incond}) reduce to 
the {\it same} Cauchy problem  {\it for all $Z$}:
\bea
&& \Delta '=\displaystyle\frac {1+v}{2s^2}\!-\!\frac 12,\qquad\quad
s'=M\Delta,\qquad M  \!:=\!\frac{4\pi e^2n_0}{mc^2}\!\equiv\!\frac{\omega_p^2}{c^2}, \label{e1} \\ 
&&  \Delta (0)\!=\!0, \qquad\qquad\qquad   s(0)\!=\! 1.\label{e2}
\eea
These are the equations of motion of a  relativistic harmonic oscillator  with a forcing term 
$v$.  In fig. \ref{graphs} we depict  the solution  corresponding to the pulse of fig. \ref{Plane}-right (with $l\!\simeq\! 27\mu$m) and  to \
$n_0\!=\! 2\! \times\! 10^{18}$cm$^{-3}$; \ $s(\xi)$ is indeed insensitive to the fast oscillations of $\Bep$ (see remark  \ref{GenRes}.\ref{insensitive}),  $\Delta(\xi)$ grows positive  for small $\xi$. The other unknowns are obtained through (\ref{hatsol}).
After the pulse is passed  the solution becomes periodic with period $\xi_{\scriptscriptstyle H}\!\simeq\!  49\mu$m. These  $l$, $n_0$ 
fulfill 
\be
2l\lesssim \xi_{{\scriptscriptstyle H}}=c t_{{\scriptscriptstyle H}},                             \label{Lncond}
\ee
where $t_{{\scriptscriptstyle H}}$  is  the plasma period associated to $n_0$ 
(recall that $t_{{\scriptscriptstyle H}}\!\ge\!t_{{\scriptscriptstyle H}}^{{\scriptscriptstyle nr}}\!\equiv\!2\pi/\omega_p
$,   the non-relativistic limit of $t_{{\scriptscriptstyle H}}$\footnote{When $\hat v\!=\!0$ then (\ref{e1}) implies $\Delta''\!=\!-M \Delta/\hat s^3$. In the nonrelativistic regime $\hat s\!\simeq\! 1$,  $\tilde \xi(t)\!\simeq\! ct$, $c d/d\xi\!\simeq\! d/dt$,
and this becomes the nonrelativistic harmonic equation $ \ddot \Delta\!=\!- \omega_p^2\Delta$.}).
For all layers of electrons with initial $Z>\Delta_{{\scriptscriptstyle M}}$ ($\Delta_{{\scriptscriptstyle M}}$ is the oscillation amplitude) it is $\hat z_e(\xi,Z)=Z\!+\!\Delta(\xi)$  for all $\xi$ (because this keeps positive for all $\xi$), $\bu(t,z)=\hat\bu(ct\!-\!z)$, and similarly for all other  Eulerian fields: a plasma wave with spacial period 
$\xi_{{\scriptscriptstyle H}}$ and phase velocity $c$  trails the pulse \cite{CatFio18,Fio17c}. On the other hand,  if $Z<\Delta_{{\scriptscriptstyle M}}$ then $\hat z_e(\xi,Z)=Z\!+\!\Delta(\xi)$  becomes
negative at some $\xi=\xi_e$, namely the  layers of electrons with such initial $Z$ exit the plasma
bulk; in the $\xi$-intervals where $Z\!+\!\Delta(\xi)\!<\!0$ the ruling equation (\ref{heq1}b) becomes \ $\hat s'_e(\xi,Z)\!=\!-MZ$.
Condition (\ref{Lncond})  secures both that the pulse is completely inside the bulk before any electron gets out of it, and that the spacial period of the plasma wave is larger that the pulse length.

Replacing these solutions  in the rhs(\ref{inteq1}) we find that $\bAp\!\simeq\!\Bap$ is indeed verified 
at least  for $t\!<\! t_c\!\simeq\!  5 \xi_{\scriptscriptstyle H}/c$. On the other hand  we find \cite{CatFio18,Fio17c}
that, while the map $z_e(t,\cdot):Z\mapsto z$ is indeed one-to-one everywhere
for $t\!<\! t_c$,  at later times wave-breaking \cite{Daw59}
(due to crossing of different $Z$-layers) occurs near the vacuum-plasma interface $Z\!\sim\!0$. This implies
that the hydrodynamic description is globally self-consistent for $t\!<\! t_c$, whereas
the use of kinetic theory  (i.e of a statistical description in phase space taking
collisions into account, e.g. by BGK \cite{BhaGroKro54} equations or effective linear inheritance relations \cite{FioMaiRen14}) is necessary if $t\!>\! t_c$, starting from a region near the vacuum-plasma interface. 
But as its effects can propagate only with a velocity smaller than $c$, they will not affect the plasma
wave trailing the pulse with phase velocity $c$.

The above  predictions  are based on idealizing the laser pulse as a plane EM wave. In a more realistic
picture the laser pulse is cylindrically symmetric around the $\vec{z}$-axis and has a {\it finite} spot radius $R$.
Using causality and heuristic arguments we can compute \cite{FioDeN16}
rough $R\!<\!\infty$ corrections to the above 
predictions: as a result, the impact of a very short and intense  laser pulse 
on the surface of a cold low-density  plasma (or gas,  ionized into a plasma by the pulse itself),
as considered e.g. in  fig. \ref{Plane}-right,  may induce [for carefully tuned $R,\widetilde{n_0}(Z) $], 
beside  a plasma traveling-wave propagating behind the pulse,
also the {\it slingshot effect} \cite{FioDeN16,FioDeN16b,FioFedDeA14},
i.e. the backward acceleration and expulsion from the plasma 
 of some surface electrons (those with smallest $Z$ and closest to  the $\vec{z}$-axis)  with remarkable energy.
For reviews see also \cite{Fio14,Fio16b,Fio18EPJ}.

\medskip
{\bf Acknowledgments.} \ The results contained in the present paper have been
partially presented in the international conference ``Wascom 2017". Devoted to Tommaso Ruggeri on the occasion of his 70th birthday.

\section{Appendix:  some useful estimates of oscillatory integrals}
\label{oscill}

Given a  function $f\in{\cal S}(\mathbb{R})$, 
integrating by parts we find  for all $n\in\mathbb{N}$
\bea
\int^\xi_{-\infty}\!\!\!\!\!\! d\zeta\: f(\zeta)e^{ik\zeta} &=& -\frac ik f(\xi)e^{ik\xi}+R_1^f(\xi) \label{modula'} \\
=\: ... &=& -\sum\limits_{h=0}^{n-1}\left(\frac ik\right)^{h+1}\!\!\! f^{(h)}\!(\xi)\,e^{ik\xi}\:+R_n^f(\xi), \qquad\mbox{where}\qquad\qquad \label{modula} 
\eea
\bea
  &&
\ba{l}
 \displaystyle R_1^f(\xi):=\frac ik \int^\xi_{-\infty}\!\!\!\!\!\! d\zeta\: f'(\zeta)\,e^{ik\zeta}=\left(\frac ik\right)^2\left[-f'(\xi)\,e^{ik\xi}
+\!\int^\xi_{-\infty}\!\!\!\!\!\! d\zeta\:
 f''(\zeta)\, e^{ik\zeta}\right] , \\[12pt]        
\displaystyle R_n^f(\xi):= \left(\!\frac ik\!\right)^n\!\!\int^\xi_{-\infty}\!\!\!\!\!\!\!\! d\zeta\:
 f^{(n)}(\zeta)\, e^{ik\zeta}=\left(\!\frac ik\!\right)^{n+1}\!\left[- f^{(n)}(\xi)\,e^{ik\xi}
+\!\int^\xi_{-\infty}\!\!\!\!\!\!\!\! d\zeta\:
  f^{(n+1)}(\zeta)\, e^{ik\zeta}\right].  
\ea        \label{Rnf}
\eea
Hence we find the following upper bounds for the remainders \ $ R_n^f$:
\bea
&& \displaystyle\left| R_1^f(\xi) \right|
\le \frac {1}{|k|^2}\left[ |f'(\xi)|+\displaystyle \int^\xi_{-\infty}\!\!\!\!\!\! d\zeta\, |f''(\zeta)|\right]
 \le \frac { \Vert f' \Vert_\infty+ \Vert f''\Vert_1}{|k|^2},
\qquad  \label{oscineqs1}\\[12pt]
&& \displaystyle\left| R_n^f(\xi) \right|
\le  \frac {1}{|k|^{n+1}}\left[  f^{(n)}(\xi)\!+\!\displaystyle \int^\xi_{-\infty}\!\!\!\!\!\! d\zeta\,|f^{(n+1)}(\zeta)|\right]
 \le \frac { \Vert f^{(n)} \Vert_\infty+ \Vert f^{(n+1)}\Vert_1}{|k|^{n+1}}
.\qquad             \label{oscineqs}
\eea
It follows $R_1^f\!=\!O(1/k^2)$, and more generally $R_n^f\!=\!O(1/k^{n+1})$, so that (\ref{modula}) are
asymptotic expansions in $1/k$.
All inequalities in (\ref{oscineqs1}-\ref{oscineqs}) are useful:
the left inequalities are more stringent, while the right ones are $\xi$-independent.

Equations (\ref{modula'}), (\ref{oscineqs1}) and $R_1^f\!=\!O(1/k^2)$ hold also if $f\in W^{2,1}(\mathbb{R})$ (a Sobolev space), in particular
if $f\in C^2(\mathbb{R})$ and $f,f',f''\in L^1(\mathbb{R})$, because the previous steps can be done also under such assumptions. Equations (\ref{modula'}) will hold with a remainder $R_1^f\!=\!O(1/k^2)$ also under weaker assumptions, 
e.g. if  $f'$ is bounded and piecewise continuous and $f,f',f''\in L^1(\mathbb{R})$, but $R_1^f$ will be
a sum of contributions like  (\ref{Rnf}) for every interval in which $f'$ is continuous.
Similarly,  (\ref{modula}), (\ref{oscineqs}) and/or $R_n^f\!=\!O(1/k^{n+1})$ hold also
under analogous weaker conditions.

\noindent
Letting $\xi\!\to\!\infty$ in (\ref{modula'}), (\ref{oscineqs1}) we find   for the Fourier  transform 
$\tilde f(k)=\displaystyle\int^{\infty}_{-\infty}\!\!\!\!\!\! d\zeta\,  f(\zeta)e^{-iky}$  of $f(\xi)$
\be
|\tilde f(k)|\le  \frac { \Vert f' \Vert_\infty+ \Vert f''\Vert_1}{|k|^2},
\ee 
hence  $\tilde f(k)=O(1/k^2)$ as well.
Actually, for functions $f\in{\cal S}(\mathbb{R})$ the decay of  $\tilde f(k)$  as $|k|\to \infty$ is 
much faster, since  $\tilde f\in{\cal S}(\mathbb{R})$ as well.
For instance, for the gaussian $f(\xi)=\exp[-\xi^2/2\sigma]$ it is $\tilde f(k)=\sqrt{\pi\sigma}\exp[-k^2\sigma/2] $.

To prove approximation (\ref{slowmodappr}) now we just need to choose $f=\epsilon$ and note that
every component of $\Bap$ will be a combination of (\ref{modula})
and  (\ref{modula})$_{k\mapsto -k}$.

\end{document}